\documentclass[%
 reprint,
 amsmath,amssymb,
 aps,
]{revtex4-2}

\usepackage{graphicx}
\usepackage{dcolumn}
\usepackage{bm}

\usepackage{romannum}
\usepackage{gensymb}
\usepackage[usenames,dvipsnames]{xcolor}

\begin{document}

\preprint{APS/123-QED}

\title{Role of Noise in Nanostructure Formation: A Theoretical Investigation of Quantum Dots and Quantum Dot Molecules}

\author{Monika Dhankhar}
 \author{Madhav Ranganathan}
 \email{madhavr@iitk.ac.in} 
 \affiliation{ Department of Chemistry, Indian Institute of Technology Kanpur, Kanpur 208016, India. } 

\date{\today}

\begin{abstract}

We theoretically model the formation of quantum dots(QDs) and quantum dot molecules(QDMs) in silicon germanium heteroepitaxy by explicitly incorporating the role of noise in a continuum theory for surface evolution in molecular beam epitaxy. Using the connection between flux and noise, we explain how changing flux can lead to a transition from QD to QDM formation, as seen in experiments. In these systems we show a dual role of noise in nanostructure growth; one where it promotes formation of QDMs via pit nucleation, and another where it curtails QDM formation due to stochastic effects.
  
\end{abstract}

\maketitle




A significant area in the nanotechnology revolution involves the understanding, control and fabrication of heteroepitaxial semiconductor nanostructures using principles of spontaneous self-assembly as opposed to top--down lithography~\cite{ESanglRMP20040}. In systems such as Si-Ge heteroepitaxy, nanostructures such as quantum dots(QDs), quantum wires, nanoridges, quantum dot molecules(QDMs) and quantum fortresses, have been observed under different molecular beam epitaxy conditions such as growth rate, film thickness and temperature~\cite{Floro_2005_PRB}. These structures are exciting due to their special physical, electronic, and optical properties and their potential for applications in micro- and optoelectronics~\cite{ESanglRMP20040, QDuZwanen20130}. Additionally, these systems and their evolution pose fundamental questions regarding the the role of different mechanisms including stochasticity. 

Amongst the different nanostructures in the heteroepitaxial Si$_x$Ge$_{1-x}$ (with $x<1$) on Si(001) system, QDs have been extensively studied both theoretically and experimentally; revealing a great amount of information about their structural transformations during the growth and subsequent evolution~\cite{EEagleshamPRL19900,Bimberg_RMP_1999,TAquaPhysRep20130}. Another, less studied nanostructure, referred to as a QDM, involves the self assembly of four QDs around a central pit\cite{KMurthy_1998_PRL,Floro_2004_PRL,Wang_2009_AdvMat,Zhinovyev_2013_PRL}. 
Both the QDs and the pit are oriented along $<100>$ directions and give a square symmetric structure because of which the QDM is sometimes referred to as a quantum fortress. Both QDs and QDMs form spontaneously in the Si$_{0.7}$Ge$_{0.3}$ on Si(001) substrate under different MBE conditions. The QDs appear at quasi--equilibrium or thermodynamic conditions whereas QDMs form under kinetic--limited conditions i.e low temperature ($<750$\degree C), high deposition rates($>0.6$ ML/s) and thick films(10-60 ML)~\cite{Floro_2005_PRB}. A closer look at QDM formation reveals that the QDs around the pit have a self-limiting structure with $ \langle 105 \rangle$ facets. 

The dynamic behavior of semiconductor nanostructures is primarily governed by deposition and surface diffusion~\cite{TSpencerPRL19910}. In unstrained(homoepitaxial) systems grown along the high symmetry direction, surface diffusion favors a flat surface, which competes with the flux which causes random undulations in the surface profile. In this case, the large diffusion length ensures that the stochasticity in the deposition flux cannot lead to kinetic roughening of the surface~\cite{B_PV_1998}. The situation is quite different in the case of strained heteroepitaxial systems, where the surface profile is not flat even in the absence of deposition. Here, strain leads to a long wavelength instability, known as the Asaro-Tiller-Grin'feld(ATG) instability causing the formation of QDs on the surface\cite{TAsaroMetaTrans19720,TGrinfeld19930}. The size of the QDs is set by the wavelength of the most unstable mode and is given by $\lambda_{\text{ATG}} = \gamma/Mm^2$ where $\gamma$ is the surface energy, $m$ is the lattice mismatch and $M$ is the Young's modulus~\cite{TSpencerActaMeta19920}. Thus the competition between the surface energy which favors nominally flat surfaces and the elastic strain which favors undulations decides the lateral size of the QDs. Under certain conditions such as large growth rates, QDMs form on the surface instead of QDs. The mechanism of formation of these QDMs and the role of increased deposition flux in the transition from QD formation to QDM formation is to be ascertained. 
A mechanism known as cooperative ridge trench formation has been used to explain QDM formation~\cite{Chiu_2007_PRB}. According to this mechanism, the nucleation of QDMs is activated by pit formation, but interrupted during annealing. Historically, there has been a lively debate on the role of pit versus QD as a strain relieving feature of heteroepitaxy~\cite{TTersoffPRL19940,Goldfarb_1997_Prl,Bouville_2004_PRB}. It is clear that the pits can lead to QDM formation as has seen in experimental and theoretical studies of QDM formation on pre-patterned substrates containing pits at predefined locations~\cite{Floro_2006_JAP, Zhang_2008_JAP, TVastolaPrb20110, Dvure_2018_JAP, Maroudas_2019, Dhankhar_2020}. 

In this study, we focus on the role of kinetic effects in QDM formation. In molecular beam epitaxy (MBE), higher flux is generally assumed to cause greater stochasticity in addition to faster growth. The correlations in beam intensity fluctuations are directly proportional to the flux~\cite{B_PV_1998}. Our studies reveal that the stochasticity can suppress formation of QDs, but enhance pit formation, which could eventually lead to QDM formation. However, in certain conditions, they could also suppress QDM formation as we show below.

The surface diffusion potential in heteroepitaxial nanostructures is a combination of a local anisotropic surface energy and a long ranged elastic interaction driven by mismatch strain~\cite{TSpencerPRL19910, TAquaPRB20100}. The contribution of the elastic interaction is calculated by solving the continuum elastic equations for an anisotropic material subject to boundary conditions at the free surface of the film, the film--substrate interface and deep in the substrate~\cite{Dixit_2017}. This is accomplished by an order by order expansion in the surface slope. As a result of this evolution, the scalar height field $h(x,y,t)$ denoting the surface profile can be shown to satisfy an evolution equation given by 

\begin{eqnarray}
\frac {\partial h}{\partial t} & = f_{\textrm{dep}} + \Delta  \Big ( -\left[1+\gamma_{n}(\mathbf{\hat n})+\gamma_{h}(h)\right]\Delta h  \nonumber \\ ~& + \left[1- \frac{1}{2}\left |\Delta h\right|^{2}\right]\frac{\partial \gamma_{h}}{\partial h}  -h_{xx}\frac{\partial^2 \gamma_{n}}{\partial h_{x}^2} -2h_{xy}\frac{\partial \gamma_{n}}{\partial h_{x} \partial h_{y}} \nonumber \\ ~& -h_{yy}\frac{\partial^2 \gamma_{n}}{\partial h_{y}^2}  -2(h_{x}h_{xx}-h_{y}h_{xy})\frac{\partial \gamma_{n}}{\partial h_{x}} \nonumber \\ ~& \left. -2(h_{x}h_{xy}+ h_{y}h_{yy})\frac{\partial \gamma_{n}}{\partial h_{y}} + {\varepsilon}'(x,y) \right ) 
\label{eq:2ndorderevoleq}
\end{eqnarray}
where $\gamma_h$ and $\gamma_n$ are the thickness dependent and anisotropic component of the surface energy respectively, and $h$ with subscript represents the appropriate derivative of the height field. ${\varepsilon}'(x,y)$ represents the elastic energy density whose calculation involves solution of elastic equations in the solid; making equation~\ref{eq:2ndorderevoleq} nonlocal. 
The units of length ($l_0$) and time ($\tau_0$) are defined as 
\begin{equation}
l_0 = \frac { \gamma^{\rm{f}} }  { [2(1+\nu)\varepsilon^{\rm{f}(0)}] }, ~~~~~~ \tau_0 = \frac{l_{0}^{4}}{\gamma^{\rm{f}}D_s}\label{eq:dimless}
\end{equation}
where $\nu$ is the Poisson ratio and $D_s$ is the surface diffusion coefficient and $\gamma_f$ is the surface energy of the flat film. The reduced elastic energy $\varepsilon'(x,y)$ is written in terms of the elastic energy density of the film $\varepsilon^{\rm{f}}(x,y)$ as
\begin{equation} {\varepsilon}'(x,y)= \frac{\varepsilon^{\rm{f}}(x,y)}{[2(1+\nu)\varepsilon^{\rm{f}(0)}]} \label{eq:Elastenergy} \end{equation}
where $\varepsilon^{\rm{f}(0)}$ is the elastic energy density of a uniformly strained film. Equation~\ref{eq:2ndorderevoleq} is numerically integrated using a fourth order Runge-Kutta algorithm~\cite{TKassamJSC20050} in Fourier space using a pseudospectral code with an integration step of 0.01$\tau_0$. 


To model the kinetic-limited environment of experimental studies, we add a noise component $f_\eta(x,y,t)$ to the deposition flux. This noise is due to intensity fluctuations in the beams used in MBE, so the usual features of thermal noise are not relevant here. We use a spatially uncorrelated Gaussian noise with mean zero with the probability distribution function given by 
\begin{equation} p(f_ \eta ) \propto \exp \left( - \frac { {f_\eta}^2} { 2 w_{\eta}^2 } \right) \label{Eq:Noise} 
\end{equation}

In typical modeling of MBE noise, the width of the distribution function $w_{\eta}$ is 
proportional to the flux~\cite{B_PV_1998} $f_{\textrm{dep}}$.  In this article, we will use $w_\eta$ as a parameter, and study its effect on the structural evolution.

We investigate the role of noise in QDM formation by carrying out simulations using 2 different initial configurations. The first corresponds to the case of a nominally flat film of some initial thickness over the flat substrate. We anticipate that, in the absence of noise, this choice will lead to QD formation for all films with thickness greater than some critical value. The second choice of initial conditions is that of a Gaussian-shaped pit-patterned film, over a flat substrate. The pattern expression is described in Ref.~\cite{Dixit_2020,Dhankhar_2020}; unlike in earlier work, we use here a patterned film on a flat substrate. This choice is expected to favor QDM formation when the thickness of the film is substantially greater than the critical thickness. 
All calculations are performed on a system of lateral dimensions $128 l_0 \times 128 l_0$ which is discretized into $128\times128$ grid points. We use mismatch strain and elastic moduli corresponding to the ${\rm Si}_{0.50}{\rm Ge}_{0.50}$ film on Si substrate, with elastic parameters given by $C_{11} =132.6$ GPa, $C_{12} =48.15$ GPa, $C_{44} =70.0$ GPa, corresponding to the average of the Si and Ge parameters.  The length $(l_0)$ and time $(\tau_0)$ units for this system are 8.8 nm and 0.25 sec respectively. The critical thickness for QD formation is equal to 0.195 $l_0$ = 6.3 ML.

\begin{figure}
    \centering
    \includegraphics[ scale = 0.16]{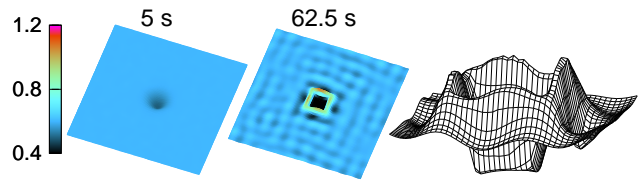}
    \caption{Height contours during evolution of patterned film on a flat substrate. The color bar on the left indicates the local film thickness in units of $l_0$ {where films thicker(thinner) than  $1.2 l_0 (0.4 l_0)$ are all uniformly colored pink (black)}. A perspective view of the QDM is shown on the rightmost panel.}
    \label{fig:Fig1}
\end{figure}

\begin{figure}
    \centering
    \includegraphics[angle = 270, scale = 0.62]{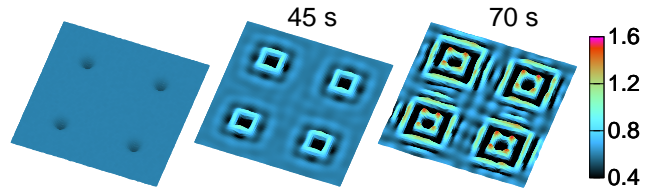}
    \caption{Snapshots showing patterned film growth on a flat substrate with 4 pits.}
    \label{fig:Fig2}
\end{figure}

We first illustrate QDM formation in simulations starting from a patterned film as illustrated in Fig.~\ref{fig:Fig1}.  We create a single Gaussian shaped pit of width $~15 l_0$ and depth $0.1 l_0$ on the film of thickness $0.4 l_0$ and study the film evolution in the absence of deposition and, consequently, noise. The QDM forms around the pit periphery and the four QDs of the QDM are connected to each other to give a fortress like structure as is clearly seen in the rightmost panel of Fig.~\ref{fig:Fig1}. When films containing multiple pits are used, the quantum fortresses become double walled as seen in Fig.~\ref{fig:Fig2}. This structure highlights the role of  interactions between the QDMs and the finite size effects which lead to interactions with the periodic images. Similar effects have been observed before for growth on patterned substrates~\cite{FLoro_2003_MSEB, Dhankhar_2020}. In this work, only the film is patterned and the substrate is flat. The film thickness and the pattern geometry is chosen appropriately in order to observe QDM formation.

\begin{figure}
    \centering
    \includegraphics[angle = 270, scale = 0.64]{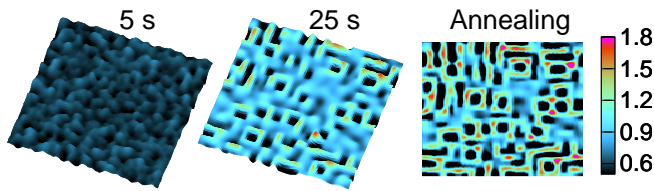}
    \caption{Growth of film on a flat substrate in presence of noise followed by annealing. For detailed description, see text}
    \label{fig:Fig3}
\end{figure}

Thus, it is clear that pre--patterned pits can serve as sites for QDM formation. This raises a fundamental question: Is it possible to observe QDM formation during growth where pits form due to stochasticity in deposition flux ? To address this question, we performed growth studies on initially flat films with deposition including stochasticity. An example of this growth is illustrated in Fig.~\ref{fig:Fig3}, where we start with an initially flat film of thickness $0.5 l_0$ with a deposition rate of $0.6$ ML/s and noise width $w_\eta = 0.15$ML. Here, we can see the formation of a surface consisting of several QDMs. The early stages of growth show undulations across the surface with pits and ridges. Some of the pits show QDM formation whereas the rest of the surface has ridges. On further deposition, the surface shows more prominent QDM formation. The third panel shows the effect of annealing the second panel where the evolution is carried out without deposition or noise. Here, the structure considerably refines and the QDMs are well formed. This is a striking result which shows how QDMs can spontaneously form during growth due to the stochasticity in the deposition and will be discussed in more detail below. However, unlike in the experiments~\cite{Floro_2004_PRL}, the QDMs do not appear to be isolated from each other, but appear as features connected by ridges. 


\begin{figure}
    \centering
    \includegraphics[scale = 0.36]{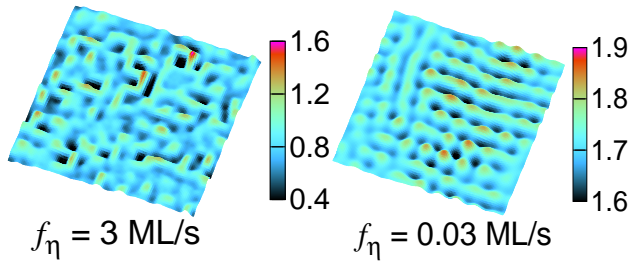}
    \caption{Snapshots during deposition at 0.6ML/s starting from a flat film of thickness $0.45 l_0$ when grown with two different noise amplitudes indicated. The left panel is after 20 seconds and the right panel is after 70 seconds. }
    \label{fig:Fig4}
\end{figure}

\begin{figure}
    \centering
    \includegraphics[angle = 270, scale =0.66]{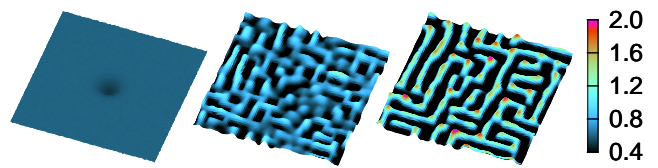}
    \caption{Snapshots showing the evolution on initial pre-patterned film of thickness $0.4 l_0$ in the presence of noise, from the initial state(left), at intermediate time(middle) and after long time (right).}
    \label{fig:Fig5}
\end{figure}

Fig.~\ref{fig:Fig4} shows the effect of noise on nanostructure formation. In the large noise case ($f_\eta=3$ ML/s), we see pit formation, but in the case of small noise ($f_\eta=0.03$ ML/s), we see an undulating surface consisting of QDs. 
Does noise always favor QDM formation over ridges or QDs ? The answer, emphatically, is {\it{no}}, as seen in Fig.~\ref{fig:Fig5}. Here we start with a configuration with a pre-patterned 0.4$l_0$ thick film and study the time evolution in the presence of noise. This system, in the absence of noise leads to QDM formation as seen in Fig.~\ref{fig:Fig1}. However, in this case, noise leads to a surface consisting of ridges as opposed to QDMs. Remarkably, the pit that was present at the beginning of the evolution is completely absent at the final configuration. It appears that, in this case, the QDM formation that would have taken place otherwise, is completely modified due to the noise. 

\par
\begin{figure}
    \centering
    \includegraphics[scale=0.37]{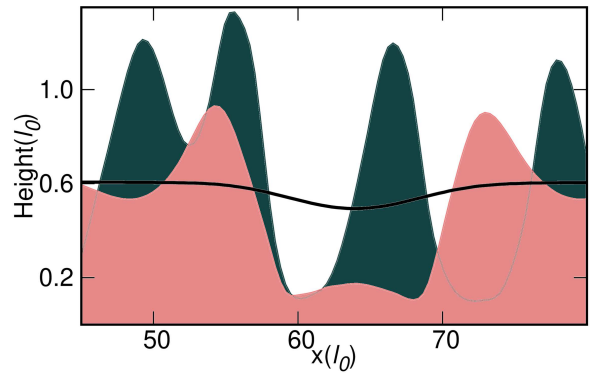}
    \caption{Evolution of profile of a film with a pit in absence of noise(pink) and in presence of noise (green). The initial surface profile(black curve) corresponds to a film with a pit at $x=64$. For details, see text.}
    \label{fig:Fig6}
\end{figure}

To get a closer look at the difference between the evolution in Figs.~\ref{fig:Fig1} and ~\ref{fig:Fig5}, we inspect the surface profile.
These cross-sectional height profiles are shown in Fig.~\ref{fig:Fig6}. We see that for the case without noise, the initial pit becomes deeper and forms QDs in the periphery, which eventually form QDMs. The presence of noise dramatically affects the surface profile evolution. In this case, the stochasticity causes breakdown of the initial pit through QD formation all over the surface. This leads to a ridged surface where the initial pit is completely subsumed. 
\par

The evolution of the pit-profile in Fig.~\ref{fig:Fig6} reveals how noise causes an existing pit to get filled to form QDs instead of the expected QDMs. On the other hand, the same noise plays a completely different role when the deposition is carried out on a flat film. For the latter case, the stochasticity due to the noise leads to the occasional pits on the surface which evolve into QDMs. Thus, noise seems to play a dual role in QDM formation. 

\par

In the absence of noise, surface evolution of a nominally flat film governed by Eq.~\ref{eq:2ndorderevoleq} shows a linear instability typical of pattern forming systems that leads to QD formation on the surface~\cite{TSpencerJAP19930}. These QDs undergo slow coarsening, which is interrupted due to anisotropy in the system~\cite{TAquaPRB20100}. The number of QDs on the surface increases during the QD formation stage and decreases during the coarsening stage. We show the effect of noise on this system. 

\begin{figure}
    \centering
    \includegraphics[angle = 270, scale= 0.65]{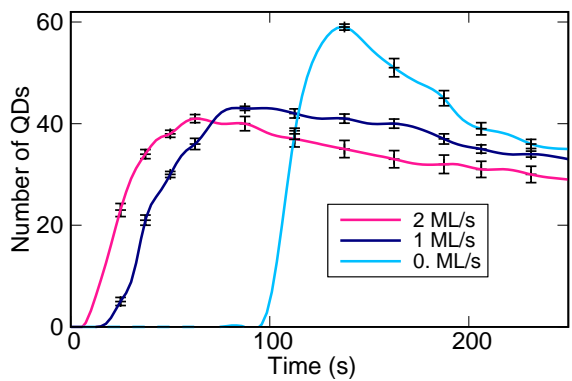}
    \caption{Change in number of QDs due to variation in noise magnitude. Initial thickness =0.25$l_0$}
    \label{fig:Fig7}
\end{figure}

The time-dependence of the total number of QDs for different noise amplitudes is plotted in Fig.~\ref{fig:Fig7}. When noise is added into the system, the initial QD formation becomes more rapid. However, this saturates at a smaller value before falling off. The final steady state value in the system with noise has smaller number of QDs than in the system without noise. This illustrates how noise initially favors QD formation but suppresses it later. This dual role of noise in QD formation is also seen in QDM formation. On a flat film, noise helps in nucleation of pits which evolve into QDMs. On a pit-patterned film, noise suppresses QDM formation. 

\par

Noise-induced pattern formation is known to be a feature of several models~\cite{Hohenberg_RMP_1993, Lindenberg_2003_PRE,Stegemann_2005_PRE, Ojalvo_2007_RevModPhys}; this work shows how noise can play an important role in nanostructure formation. We have shown the role of stochasticity due to beam fluctuations in nanostructure formation and evolution in heteroepitaxy via a comparison between the formation of QD and QDM nanostructures. 
The formation of QDM due to growth on an initially flat film is enhanced by higher growth rates, when the inhomogeneity in the deposition is significant. In this case, the stochasticity in the flux leads to formation of pits at random locations on the surface, which eventually become QDMs. For pit-patterned films, the QDMs form in the absence of noise. Adding noise tends to randomize this behavior and leads to breakdown of the QDMs. This shows how the same QDM nanostructure can either be stabilized or destabilized by the flux. Though the calculations discussed here are specific to the Si-Ge system,  the results are generic to strained heteroepitaxy; since our model is formulated in terms of dimensionless parameters. The exact details such as the size and shape of the nanostructures is expected to be different but the generic features are universal. 



\bibliography{QDM}

\end{document}